 \documentclass[preprint,review,sort&compress,12pt]{elsarticle}
\usepackage{hyperref}
\hypersetup{
    colorlinks=true,        
    linkcolor=blue,          
    citecolor=blue,        
    filecolor=magenta,      
    urlcolor=blue           
}

\usepackage{amssymb}

\usepackage{amsmath}

\usepackage[modulo]{lineno}

\journal{Nuclear Instruments and Methods in Physics Research A}

\begin{document}

\begin{frontmatter}



\title{Comparative Analysis of Electron Acceleration by Laser Pulse in Flat and Chip Dielectric Structures}


\author{G.V. Sotnikov\corref{cor1}}
\ead{sotnikov@kipt.kharkov.ua}
\cortext[cor1]{Corresponding author}

\author{A.V. Vasiliev}

\author{I.V.~Beznosenko}

\author{S.M.~Kovalov}

\author{A.I.~Povrozin}

\author{O.O.~Svystunov}

\address{National Science Center Kharkiv Institute of Physics and Technology %
1, Akademichna St., Kharkiv, 61108, Ukraine}

\begin{abstract}
A comparative analysis of two types of dielectric laser accelerators (DLA) based on periodic (grating) and flat dielectric structures to accelerate electrons in the energy range from $300~keV$ to $3\,GeV$ is presented. The main attention is paid to the conditions, efficiency and restrictions of each acceleration method, as well as the influence of laser radiation parameters on electron acceleration processes. Single and double (both grating and flat) dielectric structures and their impact on acceleration are considered. For the study of two types of quartz DLA, the Ti:Sa laser system with a generation band width $790-810~nm$ (FWHM), the laser electric field $6~GeV/m$  are used. The study showed that a flat dielectric structure provides more effective acceleration in a wide range of energies, especially with a symmetrical geometry (double structures), compared with the periodic structure. If we consider only a periodic structure, then with the selected symmetrical geometry, for the ultra relativistic electrons, it demonstrates the acceleration rate two times of magnitude more than for single configuration. However, the use of a one-sided periodic structure turns out to be preferable for accelerating electrons with moderate energies, $\sim 0.5-0.9~MeV$, where the acceleration rate in a one-sided configuration is higher than in a symmetric (double) periodic structure. The space-time distributions of laser-excited electromagnetic fields in the accelerating channel and their influence on the electron beam is analyzed also. The advantage of a flat structure over a periodic one, which arises due to the design features of the corresponding dielectric accelerators, is discussed.
\end{abstract}




\end{frontmatter}


\section{Introduction}
To use dielectric structures for acceleration by waves arising from their illumination with laser beams was proposed 60 years ago~\cite{Shimoda1962AO,Lohmann1962TN}. These perspective devices were called dielectric laser accelerators (DLAs). Subsequent investigations showed the possibility of creating on the basis of proposed idea  compact accelerators with a acceleration rate of  $\sim 1~GeV/m$~\cite{Kheifets:1971zza,Nagorsky:1983_12HEACC,Lawson:1984wv,Borovsky1987PAST-3,Borovsky1987PAST-4,Fernow1991BNL}. From the very beginning, two directions in DLA researches emerged. The first direction used a smooth (flat) dielectric structures irradiated by laser beam under an angle greater full internal reflection (so called inverse Cherenkov DLA)~\cite{Shimoda1962AO,Lohmann1962TN,Kheifets:1971zza,Nagorsky:1983_12HEACC,Lawson:1984wv,Fernow1991BNL}.  The second direction proposed to use grating (chip) dielectric structures(so called inverse Smith-Purcell DLA)~~\cite{Lohmann1962TN,Lawson:1984wv,Borovsky1987PAST-3,Borovsky1987PAST-4}.

Interest in the use of laser beams to accelerate charged particles has renewed in the last decade due to the widespread use of lasers at the TW level of the micron wavelength and pulse durations 100~fs and less. This progress in laser technology and the development of dielectric materials with breakdown voltages significantly higher than in all-metal structures~\cite{Lenzner1998PRL,Bendib2013LPB,Soong2012AIP,Hobbs2012SPIE} have made it possible to proceed to experimental verification and implementation of the basic principles of laser accelerators based on dielectric structures (see~\cite{Peralta2013Nature,Breur2013PRL,England2014RMP,Wootton2016,Sun2021NJP,England_2022, Niedermayer_2022} and reference there).

The above two historically established directions in the development of dielectric laser accelerators are still developing independently of each other. Both directions study ways to increase the efficiency of acceleration~\cite{Zhang2023TPS,PhysRevAccelBeams.27.051303}, the stability of accelerated particle bunches, etc. At the same time, there is no comparative analysis  of these two methods in the literature. The simplest comparison is the acceleration rates in flat and chip dielectric structures with the same parameters of laser pulses.

In this paper we present the results of comparative analysis of electron beams acceleration by laser pulse in flat and chip DLAs. For this analysis we will consider both single and double dielectric structures.  The paper is organized as follows. The statement of problem is formulated in Section~\ref{sec:2}. Here we present dielectric structures chosen for the analyses and give parameters of laser pulse and electron beams used in the next numerical simulations. Section~\ref{sec:3} deals with numerical analysis of electron beam acceleration. Acceleration rates in dependance from initial energy are determined.  The difference in acceleration rates in flat and chip structures is discussed. Section~\ref{sec:Conclusion} gives the summary.

\section{Statement of the problem}
\label{sec:2}
\begin{figure}
    \includegraphics[width=0.47\textwidth]{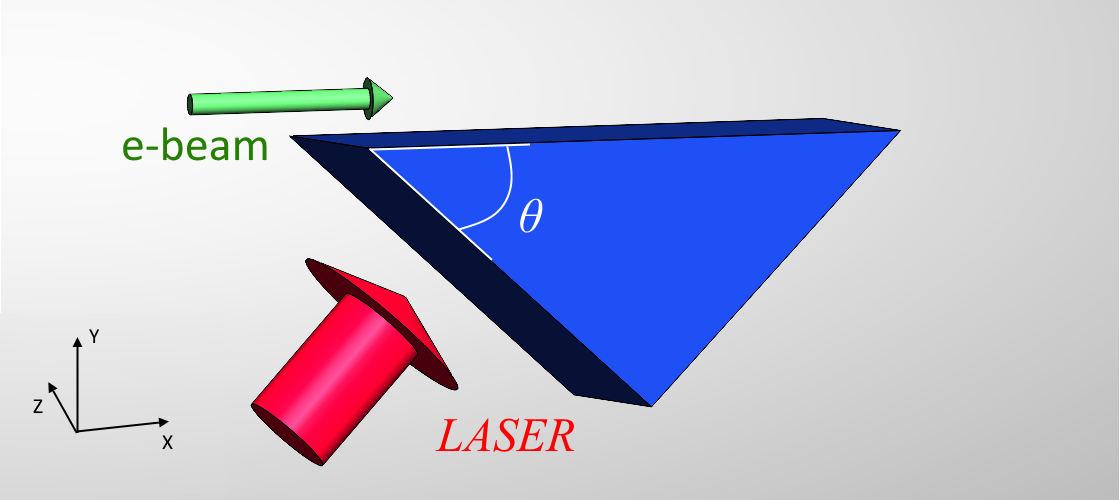}\hfil \includegraphics[width=0.47\textwidth]{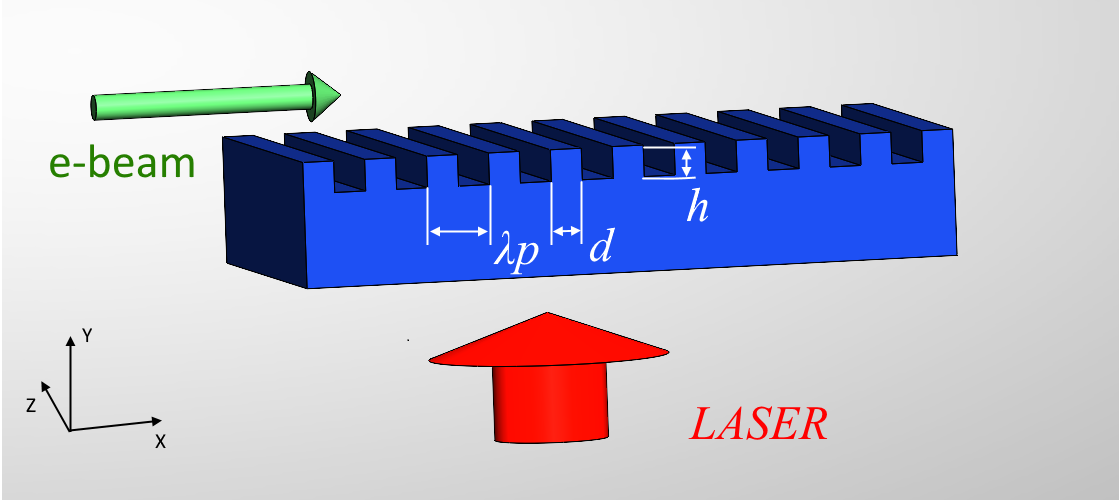}\\
    \includegraphics[width=0.47\textwidth]{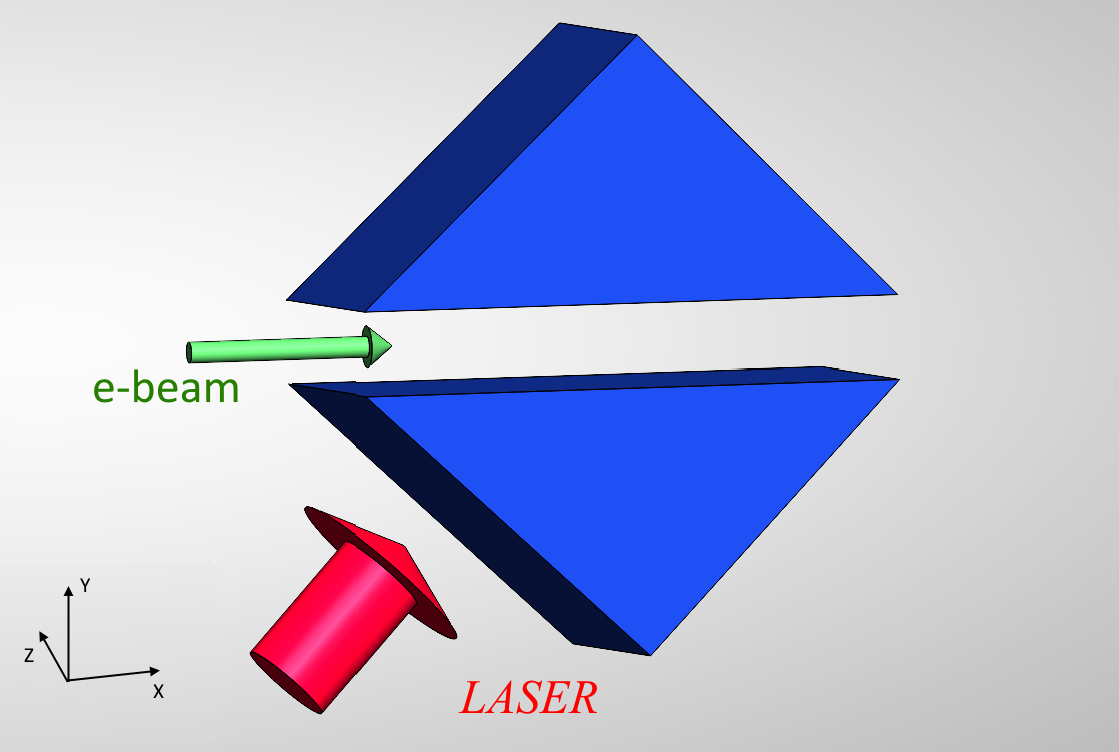}\hfil \includegraphics[width=0.47\textwidth]{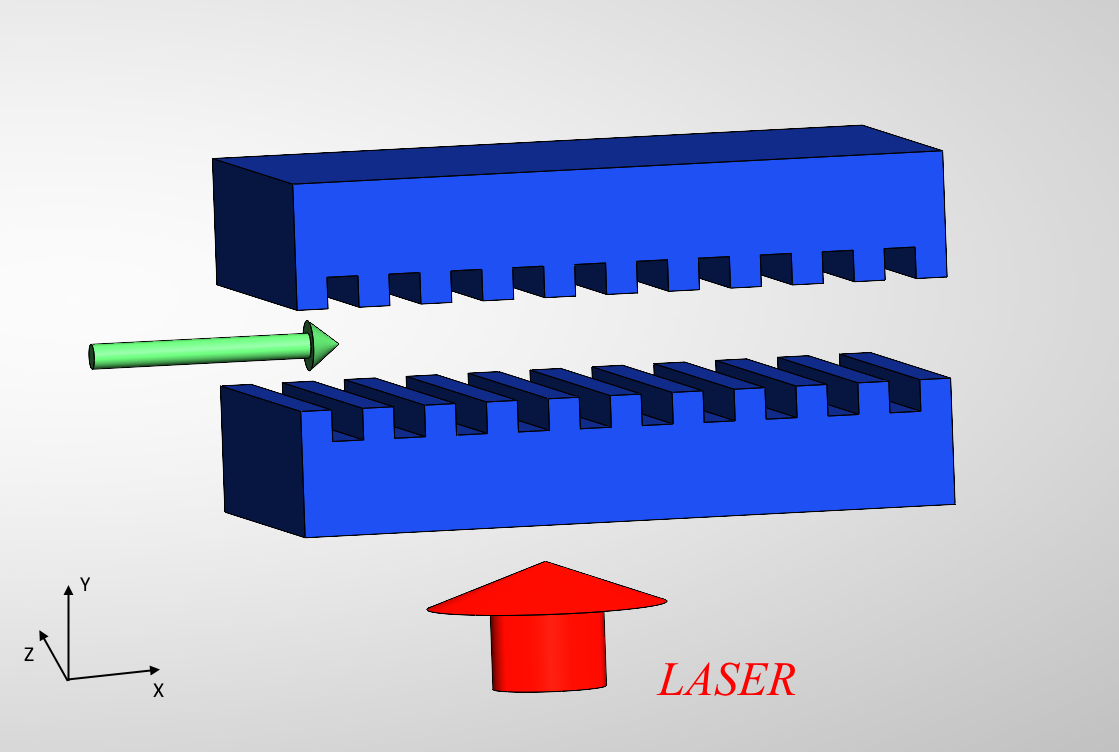}\\
    \caption{Schematic view of dielectric laser accelerating structures under investigation. Directions of injection electron beam  and laser beam are shown relative to chosen coordinate system.}\label{Fig:01}
  \end{figure}

The accelerating structure under consideration are shown in Fig.\ref{Fig:01}. The left column shows flat dielectric structures, the right column shows chip dielectric structures. The top row shows single structures, or asymmetric, and the bottom row shows double structures, or symmetric.

The laser beam falls from free space onto the lateral surface of the dielectric structure at a right angle. The prism angle for flat structures is determined from the condition of ensuring Cherenkov synchronism between the beam and the evanescent wave arising at the dilectric-vacuum interface,
\begin{equation}\label{eq:01}
  \sin(\theta)=1/\beta_0\sqrt{\varepsilon},
\end{equation}
$\beta_0=v_0/c$, $v_0$ is an initial velocity of beam electrons, $\varepsilon$ is a permittivity of dielctric material,  $c$ is the speed of light in free space. For ultrarelativistic electrons   $\beta_0 \to 1$  the prism angle $\theta$  is equal to the limit angle of full internal reflection.

The period of the dielectric chip structures is also determined from the condition of maintaining synchronism between the electron and the spatial harmonic of the wave that occurs when a laser beam falls on the periodic structure~\cite{Palmer1980PA}
\begin{equation}\label{eq:02}
  \lambda_p=n\beta_0\lambda_L,
\end{equation}
$n$ is the number of the resonant harmonic.

In the case of double structures, we used unilateral irradiation of the dielectric structures. Such unilateral irradiation leads to the appearance of an asymmetric part in the amplitude of the accelerating field for both flat structures~\cite{Zhang_2023, Vasiliev2023PAST} and chip structures~\cite{Breuer_2014_JPB}, for ultrarelativistic electron beams such a term tends to zero~\cite{Vasiliev2023PAST}. The electromagnetic field of the laser pulse has p-polarization, the electric field lies in the $X-Y$ plane of Fig.~\ref{Fig:01} and has the gauss form:
\begin{multline}\label{eq:03}
  E_s(r,s,t) = E_L\frac{w_0}{w(s)}\exp\left[-\frac{w_0^2}{w^2(s)}\right]\exp\left[-2\ln(2)\frac{(s-ct)^2}{c^2\tau_0^2}\right]\\
  \times \mathfrak{R}\left\{\exp[i\omega_0 t -ik_0s - ik_0\frac{r^2}{2R(s)}+i\psi_g(s)]\right\},
\end{multline}
where $E_p$ is an amplitude of the electric field, $s$ is  a distance along a laser propagation direction,  $r$ is transverse to laser propagation direction coordinate, $w_0$ is a waist or the smallest transverse size of the laser in the focal plane ($s=0$); $\tau_0$ is full width at half power maximum of the pulse duration; $k_0 = 2\pi/\lambda_L$ and $\omega_0=ck_0$  represent the wave number and angular frequency of the laser beam with the wavelength $\lambda_L$ respectively, $w(x)=w_0\sqrt{1+(s/s_R)^2}$ is the waist at distance $s$ from the focal plane,  $s_R=\pi w_0^2/\lambda_L$  is the Rayleigh length, $R(s)=  s[1+(s_R/s)^2]$ is the  radius of curvature of the wave front, and $\psi_g (s)= arctan(s/s_R)$ is the Gouy phase shift as the function of propagation distance $s$.

A monoenergetic electron bunch with a uniform transverse density distribution and a Gaussian longitudinal density distribution is injected along the surface of the dielectric structures at some distance from them.

\section{Numerical analysis}\label{sec:3}

For the simulations we used the PIC solver of the CST PARTICLE STUDIO code of the CST STUDIO SUITE~\cite{CST-Particle}. An electromagnetic field of the Gaussian laser beam (eq.~\ref{eq:03}) was determined by an imported VBA macro embedded in the numerical code. The laser beam focus was at the dielectric-accelerator channel boundary. Due to the Gaussian shape of the laser pulse in space and time (see eq.~\ref{eq:03}), for efficient acceleration of the electron beam in periodic chip structures and flat structures, it is necessary to precisely synchronize the electron beam with the laser pulse. In the numerical simulation, this was achieved as follows. In the case of a periodic chip structure, the beam was synchronized with the laser radiation by delaying the electron beam. The delay is chosen so that the electron beam reaches the center of the focal spot (waist) of the laser radiation exactly at the moment when the laser radiation reaches its maximum in the acceleration channel. This ensures maximum acceleration of electrons, since the peak of the laser pulse and the peak of the electron beam coincide in space and time, which allows the most efficient use of the laser energy to accelerate electrons.
For a flat structure, where the accelerating phase moves with a velocity close to the velocity of the beam electrons, the electron beam (its center) was injected directly into the maximum of the accelerating phase of the laser radiation. A group of electrons near the maximum of the accelerating phase will interact with the electromagnetic field for the longest time, which will provide the maximum energy gain.

In our numerical simulations, we applied the following boundary conditions. The X-boundaries are set as electric boundary conditions (the tangential component of the electric field is zero). The Y-boundaries are set as open free space to eliminate the influence of the reflected wave on the beam electrons. The Z-boundaries are set as magnetic boundary conditions (the tangential component of the magnetic field is zero).

Table~\ref{Tbl:01} shows the parameters of the dielectric structures, laser beam and electron beam used in the numerical simulation. The laser beam~\cite{Vasiliev2018PAST} had a central wavelength of $800~nm$ with a bandwidth of 20 nm, the electric field strength of the wave was chosen to be $6\cdot 10^9~V/m$, which is close to the threshold value of fused silica breakdown~\cite{Breur2014PRSTAB}. There are extensive studies in the literature on the ratio of the width and height of pillars for chip structures with a fixed period to ensure the expected acceleration gradients~~\cite{Plettner2006PRSTAB,Peralta_PhD,Wei_PhD}. For simplicity, we limited ourselves to the case when the height and width of the pillars are equal to half the period of the structure $h_p=w_p=\lambda_p/2$.
\begin{table}
\centering
\caption{Parameters of flat and chip dielectric structures, laser pulse and electron beam for PIC simulation of DLAs}\label{Tbl:01}   
\vspace{6pt}
\begin{tabular}{l c c}              
\hline\hline                        
\textbf{Item} & \textbf{Value} & \textbf{Units} \\ [0.5ex] 
\hline\hline                              
Central wavelength of pulse $\lambda_L$ & 800 & nm \\
Laser spectrum width   & 20 & nm \\
Laser pulse waist  $w_0$ & 10 & $\mu m$ \\
Wave amplitude  $E_L$ & $6\cdot 10^9$ &  V/m \\
FWHM duration of Laser pulse  $\tau_0$ & 160 &  fs  \\
Dielectric permittivity of structure $\varepsilon$ & 2.112 & fused silica \\
Period of chip structure $\lambda_p=\beta_0\lambda_L$ & $ \leq 800$ & nm \\
Pillar height $h_p$ and width $w_p$ of chip structure, $\lambda_p/2$ & $\leq 400$ & nm \\
Limit angle of the full internal reflection $\vartheta $ &  43.48  & degree \\
Electron beam energy $W_b$ & $0.3\div 3000$ & MeV\\
Electron beam transverse size &  50 & nm \\
Vertical distance of beam from surface & 200 & nm \\
Vacuum gap for symmetrical case & 400 & nm \\
FWHM electron bunch duration, $\tau_b$   &  2 & fs \\
Full electron bunch duration  &  6  & fs \\
\hline\hline                                  
\end{tabular}
\end{table}

\begin{figure}
\centering
\includegraphics[width=0.75\textwidth]{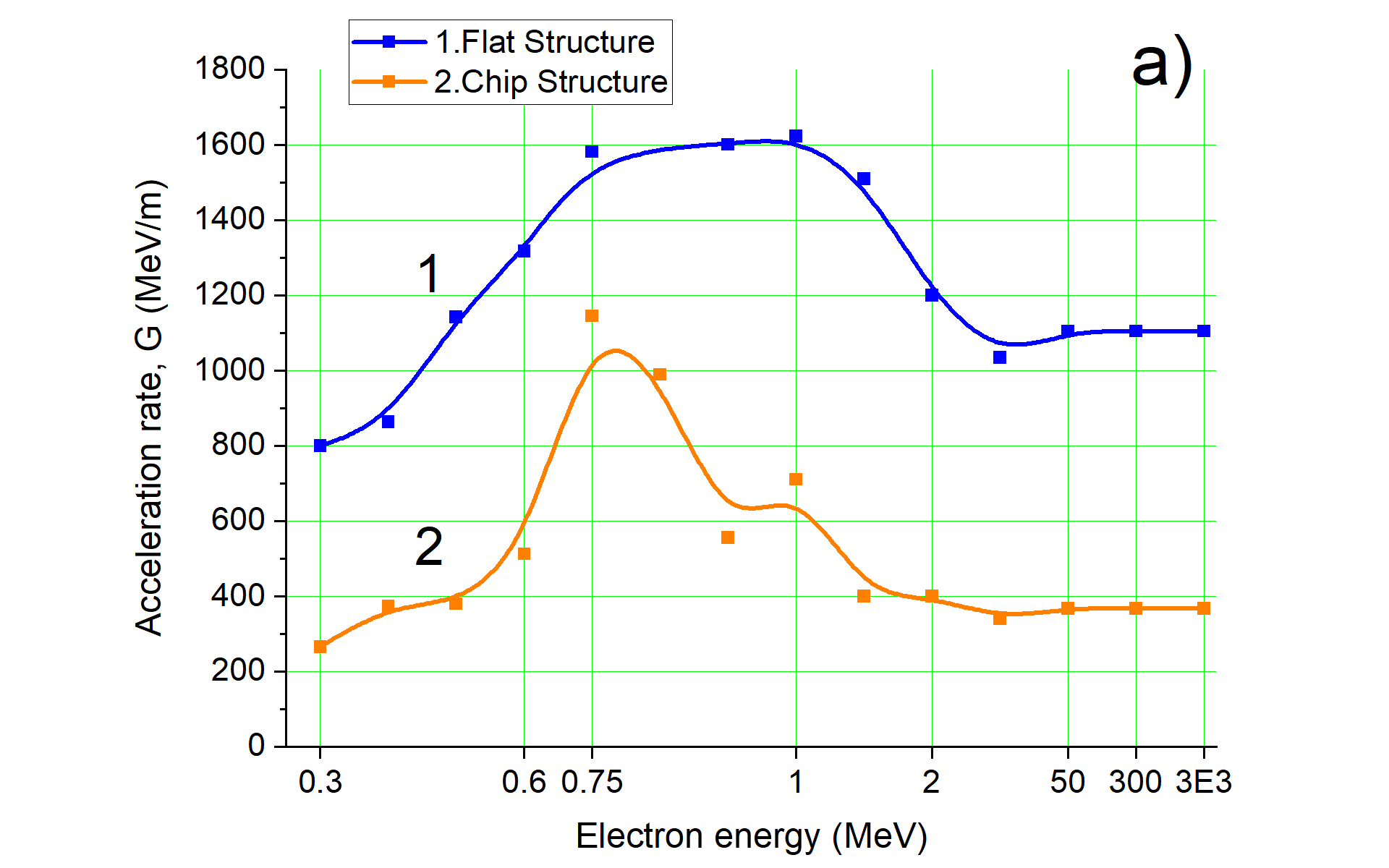}\\
\includegraphics[width=0.75\textwidth]{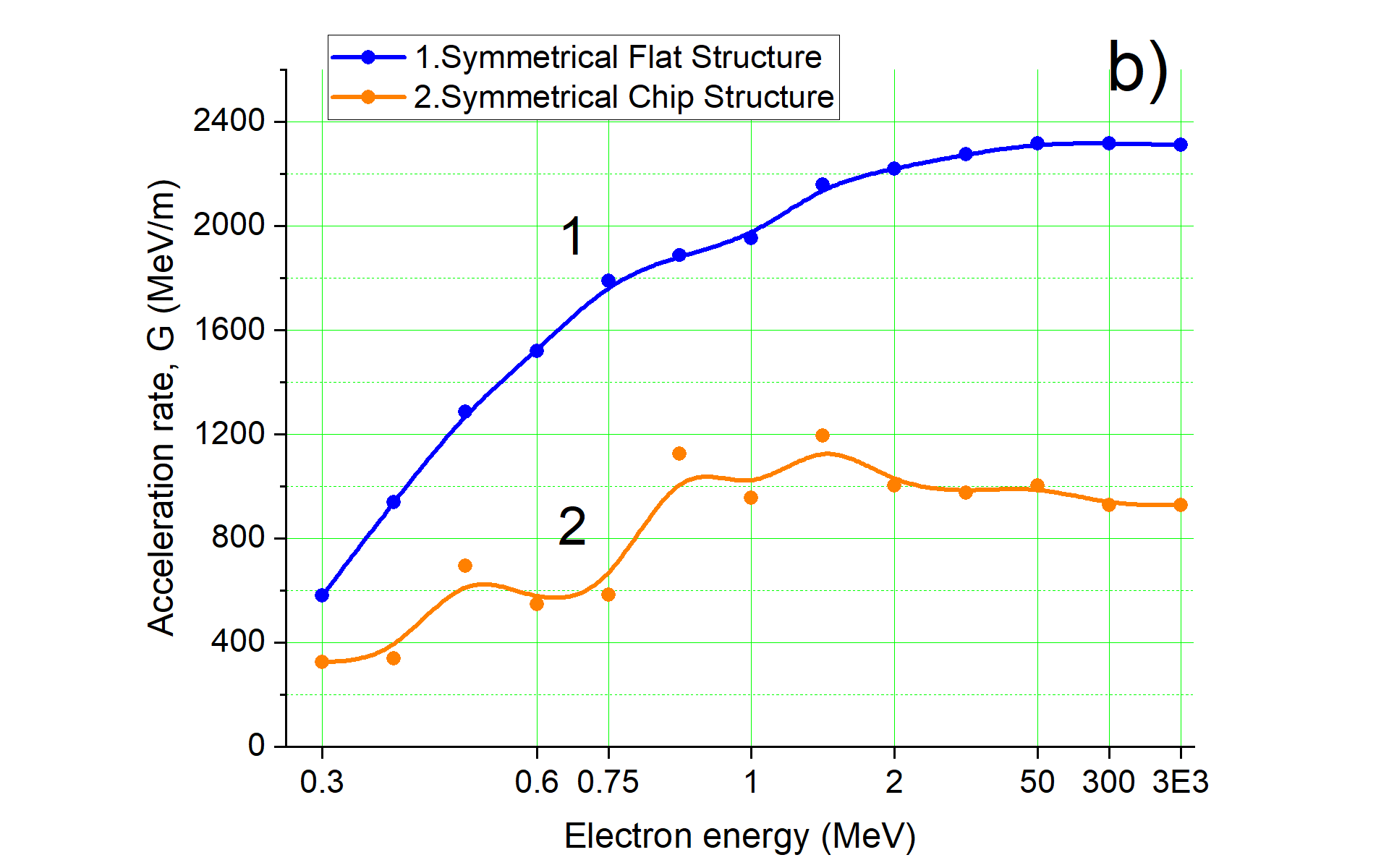}\\
\caption{Acceleration rate versus the initial energy of the electron beam in: a) single dielectric structures, b) dielectric double (symmetrical) structures. Acceleration rates are computed using maximally accelerated  electrons.}\label{Fig:02}
\end{figure}

Fig.\ref{Fig:02} shows the results of the simulations the acceleration of an electron beam in a dielectric laser accelerator of flat and chip configurations. Fig.\ref{Fig:02}a shows the acceleration rate depending on the initial energy of the electron beam in single structures, Fig.\ref{Fig:02}b shows the acceleration rate in double, or symmetric, structures. The acceleration rate shown in these figures were calculated using the maximally accelerated electrons at a distance of ~20 microns.
We started with an initial beam electron energy of 300~keV (the prism angle $\theta$ of the flat fused silica structure is 63 degrees), further reduction of the initial energy significantly increases the prism angle and the unusable length of the flat accelerator structure).

For a single (nonsymmetric) configuration of accelerator structures, with an increase in the initial electron energy from 300~keV, the acceleration rate increases for both flat and chip structures. This is due to an increase in the amplitude of the transmitted wave with an increase in its phase velocity \cite{Kozak:17,Bolshov2021PAST,Zhang_2023}. Then it reaches a maximum at 750-1000~keV and with a further increase in the initial energy the acceleration rate begins to decrease due to a significant change in the polarization of the transmitted wave~\cite{Bolshov2021PAST},  then reaches saturation at energies greater than 5~MeV. The latter behavior differs from the analytical results, which at $\beta_0\to 1$ give a zero acceleration rate~\cite{Kozak:17,Bolshov2021PAST,Zhang_2023}. The possible discrepancy is due to different statements of the problems. In contrast to the current statement, in the studies on the flat structures~\cite{Kozak:17,Bolshov2021PAST,Zhang_2023} the laser pulse with a finite frequency band was replaced by a plane wave with a fixed frequency. In addition, when exciting the chip structure, even in the approximation of a plane wave with a fixed frequency, spatial harmonics of the field can be excited, which will contribute to the accelerating field.

For a double (symmetric) configuration of accelerator structures, the initial behavior of the acceleration rate is the same as for single structures, i.e. it increases with increasing beam electron energy from 300~keV. Then it reaches saturation at starting electron energies close to their values in the case of using single structures. After reaching saturation, the acceleration rate changes practically not, with the exception of its insignificant decrease in the case of chip structures at energies greater than 2~MeV. The absence of a decrease in the acceleration rate in symmetric flat structures for relativistic values of electron energy is consistent with analytical results~\cite{Frandsen2006LP,Vasiliev2023PAST,Zhang_2023}. It should be noted that the acceleration rate for relativistic electrons in symmetric dielectric structures, both flat and chip, is significantly higher than in single structures.

\begin{figure}
\centering
\includegraphics[width=0.75\textwidth]{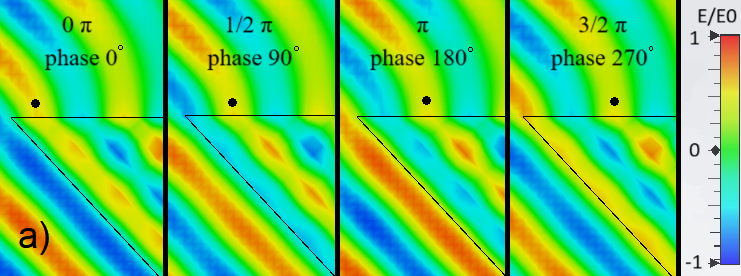}\\
 \includegraphics[width=0.75\textwidth]{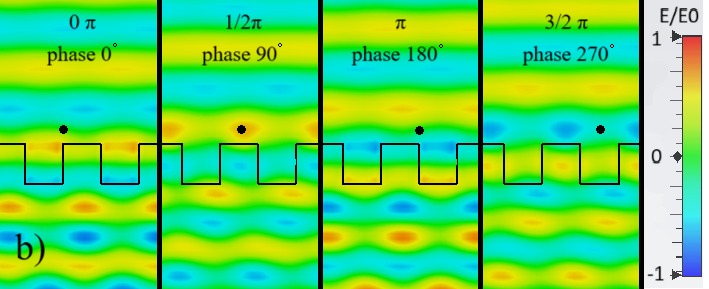}\\
\caption{Snapshots of the axial electric field distribution and the position of accelerated resonant electrons (black circles) for four consecutive moments of time shifted by a quarter of the period of the excited wave. The figure~\ref{Fig:03}a  is given for a single flat dielectric structure, the figure~\ref{Fig:03}b is given for a single chip structure. Electron energy is 1 MeV. }\label{Fig:03}
\end{figure}

\begin{figure}
\centering
\includegraphics[width=0.7\textwidth]{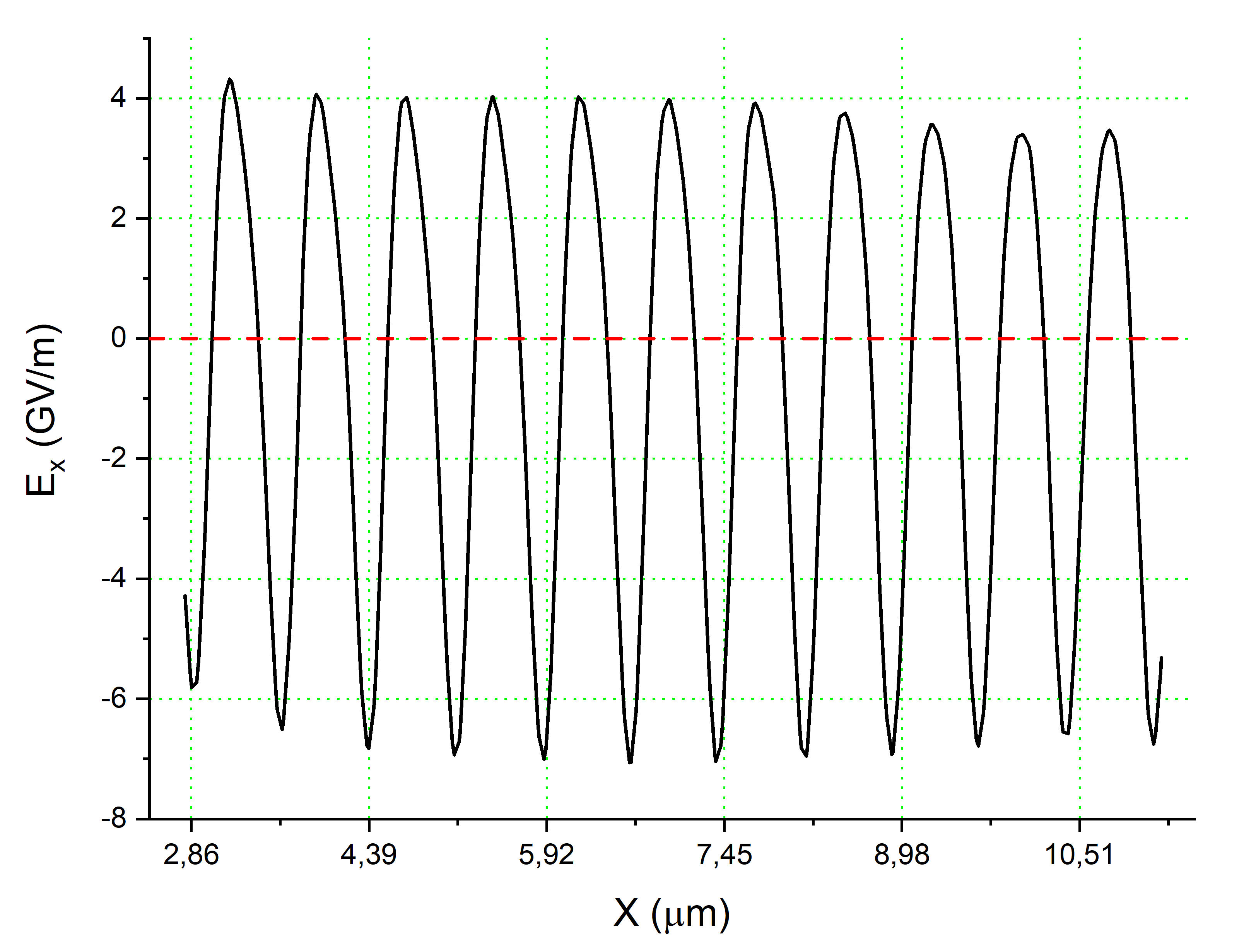}\\
\caption{Axial electric field acting on resonant electron during its acceleration in single chip structure. The initial beam energy $W_b=1~MeV$. The vertical dashed lines are plotted in multiples of two periods of the structure.}\label{Fig:06}
\end{figure}

From Fig.\ref{Fig:02}a and Fig.\ref{Fig:02}b one can see that the use of flat structures provides a significantly higher acceleration rate than chip structures (by 1.5-3 times). Such a significant difference can be explained by different physical mechanisms underlying the acceleration of electrons in these accelerating structures. In flat dielectric structures, resonant acceleration occurs due to the inverse Cherenkov effect (equation~\ref{eq:01}). Figure~\ref{Fig:03}a shows the maps of the axial field distribution and the location of accelerated point particles for different times, shifted by a quarter of the wave period.
One can see that throughout the entire wave period, the electron is in the accelerating phase of the wave, and this will continue until it leaves the resonance with the wave. Figure~\ref{Fig:03}b shows maps of the axial field distribution and the location of accelerated point particles similar to Fig.~\ref{Fig:03}a in the case of a chip structure. In chip dielectric structures, resonant acceleration occurs due to the inverse Smith-Purcell effect. It can be seen from Fig.~\ref{Fig:03}b that the longitudinal field acting on the resonant particle (equation~\ref{eq:01}) changes sign to the opposite on one period of the structure. Since we consider acceleration at the first spatial harmonic ($n=1$), such a change occurs once~\cite{Breur2013PRL}. The electron is subject to an accelerating field when it is above the pillars and a decelerating field when it moves above the grooves of the structure. Due to the excess of the amplitude of the accelerating field over the decelerating field, the electron as a whole is accelerated. The excess of the accelerating field over the decelerating field, along with Fig.~\ref{Fig:03}, is clearly demonstrated by Fig.~\ref{Fig:06}, which shows the longitudinal electric field acting on the resonant electron during its acceleration in a single chip structure. It is evident that the average value of the accelerating field (negative value) exceeds its decelerating value.

\begin{figure}
\centering
\includegraphics[width=0.7\textwidth]{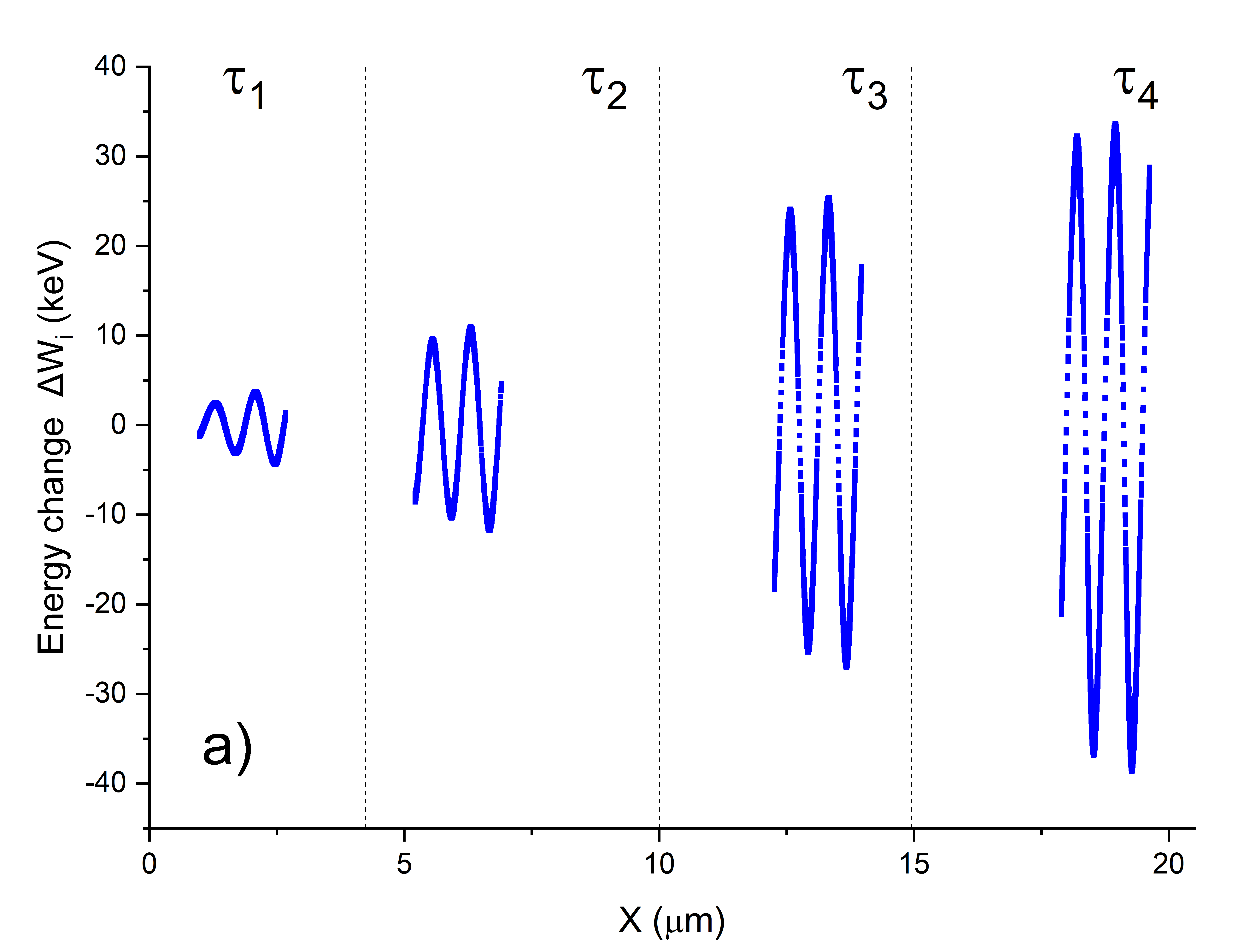}\\
\includegraphics[width=0.7\textwidth]{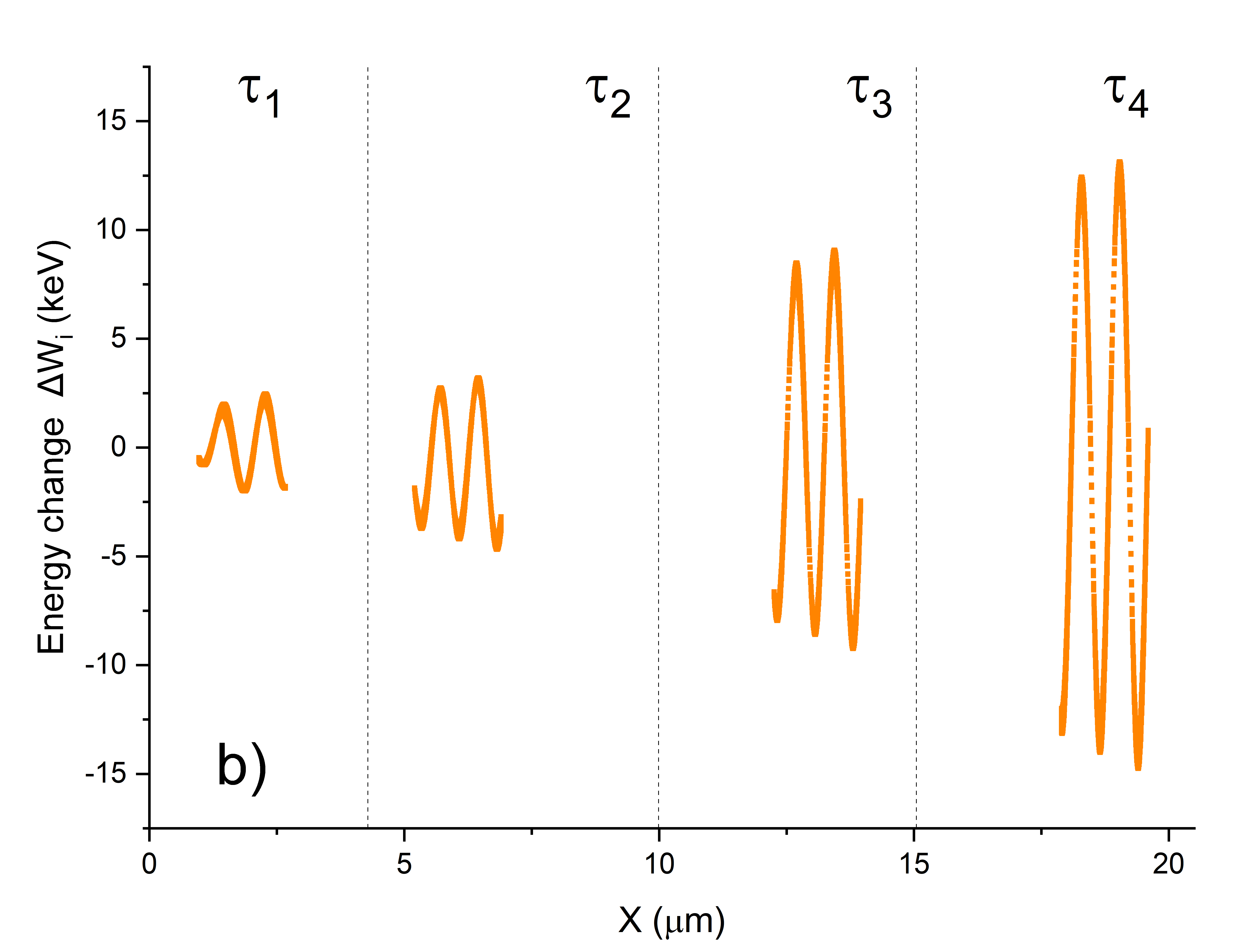}\\
\caption{Dynamics of beam electron energy distribution  during its transport in a flat DLA(a) and chip DLA(b). Here Energy change $\Delta W_i=W_i-W_b$, $W_i$ is current energy of i-th electron. The initial beam energy $W_b=1~MeV$.}\label{Fig:04}
\end{figure}

The dynamics of the electron acceleration of an extended beam in the studied single dielectric structures is shown in Fig. ~\ref{Fig:04}.
Figure~\ref{Fig:04}a describes the change in the electron energy in a flat DLA, and Fig.~\ref{Fig:04}b in a chip DLA. The beam length is equal to the double central wavelength of the laser pulse. The initial electron energy is 1~MeV. The energy change is shown for four time moments shifted by a quarter of the time period of the wave along the accelerating unit. It can be seen that, in accordance with the previous figure, in a flat DLA we have a monotonic acceleration and deceleration of electrons (depending on the resonant phase of the wave) in contrast to the chip DLA.

\begin{figure}
\centering
\includegraphics[width=0.7\textwidth]{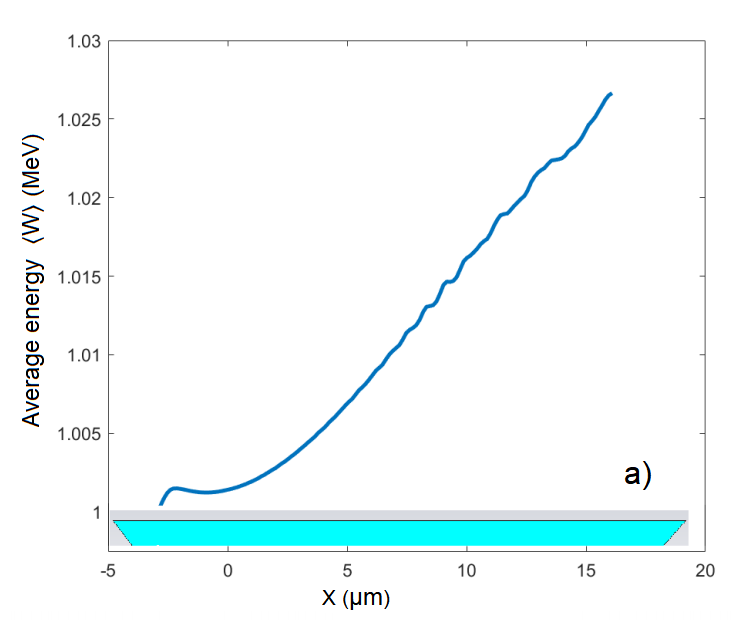}\\
 \includegraphics[width=0.7\textwidth]{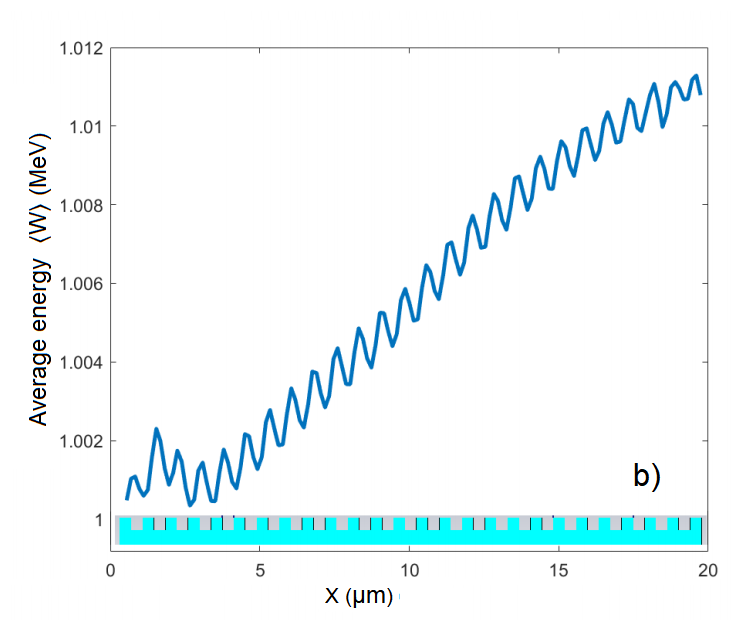}\\
\caption{Average energy of short electron beams ($\tau_b=0.1~fs$) in a flat DLA(a) and a chip DLA(b). Here Average energy $\langle W \rangle=\langle W_i\rangle$, $W_i$ is current energy of i-th electron, brackets show averaging over all electrons in the beam.  At the initial sections of the structures, one can see matching the electron beam with the wave. The initial beam energy $W_b=1~MeV$.}\label{Fig:05}
\end{figure}

In order to obtain more clear picture of electron acceleration we took very short beam, FWHM electron bunch duration is 0.1~fs, and injected in acceleration phase of the wave. So we remove decelerated particles. The results are shown in~\ref{Fig:05}.
 Figure~\ref{Fig:05}a presents average energy of electron bunch during an acceleration in flat DLA, figure~\ref{Fig:05}b shows average energy of electron bunch during an acceleration in chip DLA. These pictures confirm we said above. We have monotonic acceleration in flat DLA (except for a small initial section where the beam is matched to the excited wave);  we have nonmononotic (oscillating) total acceleration in chip DLA, where on one half of the wave period the electrons accelerate, on the other half of the wave period they decelerate, and the final balance of energy gain  on one period of the structure is positive. We obtained qualitatively similar energy gain dependences for symmetrical flat and chip structures. For chip structures, the overall increase in the average beam energy is accompanied by a deeper energy modulation on one period of the structure.

\section{Conclusions}\label{sec:Conclusion}

In conclusion, we compared the acceleration rates of electron beams in DLAs based on laser pulse excitation of flat dielectric structures (inverse Cherenkov effect) and chip dielectric structures (inverse Smith-Purcell effect).
Flat dielectric structure provides more effective acceleration in a wide range of energies, especially with a symmetrical geometry (double structures), compared with the periodic structure.
The periodic structure with symmetric geometry has an acceleration rate twice that of the single configuration.
The use periodic structure with one-sided configuration turns out to be preferable for accelerating electrons with moderate energies, less 1 MeV, where the acceleration rate is higher than in a symmetric (double) configuration.

In our analysis, we limited ourselves to comparing flat and chip structures only by a single criterion - acceleration rate. For the final choice of one or the other structure as an accelerator model of DLA, additional studies of the parameters of accelerated bunches are required.

\section*{Acknowledgments}
The study is supported by the National Research Foundation of Ukraine under the program “Excellent Science in Ukraine” (project \# 2023.03/0182).

\bibliographystyle{elsarticle-num}
\bibliography{Bibliography_DLA}

\begin{thebibliography}{10}
\expandafter\ifx\csname url\endcsname\relax
  \def\url#1{\texttt{#1}}\fi
\expandafter\ifx\csname urlprefix\endcsname\relax\def\urlprefix{URL }\fi
\expandafter\ifx\csname href\endcsname\relax
  \def\href#1#2{#2} \def\path#1{#1}\fi

\bibitem{Shimoda1962AO}
K.~Shimoda, \href{https://doi.org/10.1364/AO.1.000033}{Proposal for an electron
  accelerator using an optical maser}, Applied Optics 1 (1962) 33--35.
\newblock \href {https://doi.org/10.1364/AO.1.000033}
  {\path{doi:10.1364/AO.1.000033}}.
\newline\urlprefix\url{https://doi.org/10.1364/AO.1.000033}

\bibitem{Lohmann1962TN}
A.~Lohmann,
  \href{https://www.laserphysik.nat.fau.de/files/2020/09/lohmann_electronaccelerationbylightwaves_ibm-technote_1962.pdf}{Electron
  acceleration by light waves}, IBM Technical Note TN5 (1962) 169--182.
\newline\urlprefix\url{https://www.laserphysik.nat.fau.de/files/2020/09/lohmann_electronaccelerationbylightwaves_ibm-technote_1962.pdf}

\bibitem{Kheifets:1971zza}
S.~A. Kheifets, \href{https://inspirehep.net/literature/73386}{Motion near a
  surface on which total internal reflection of electromagnetic waves takes
  place}, in: Blewett, M.~Hildred, N.~Vogt-Nilsen (Eds.), Proceedings of 8th
  International Conference on High-Energy Accelerators(HEACC 71), Vol. C710920,
  Geneva, CERN, 1971, pp. 597--599.
\newline\urlprefix\url{https://inspirehep.net/literature/73386}

\bibitem{Nagorsky:1983_12HEACC}
G.~Nagorsky, A.~Amatuni, W.~Arutiunian,
  \href{https://inspirehep.net/literature/263523}{Resonance acceleration of
  charged particles by a surface wave arising at total internal reflection},
  in: F.~Cole(Fermilab), R.~Donaldson(Fermilab) (Eds.), Proceedings of 12th
  International Conference on High-Energy Accelerators(HEACC 83), Vol. C830811,
  Batavia, Fermilab, 1983, pp. 488--490.
\newline\urlprefix\url{https://inspirehep.net/literature/263523}

\bibitem{Lawson:1984wv}
J.~Lawson,
  \href{https://inis.iaea.org/collection/NCLCollectionStore/_Public/16/085/16085644.pdf}{Laser
  accelerators: where do we stand?}, in: Proceedings of Workshop on the
  Generation of High Fields for Particle Acceleration to Very High Energies,
  Frascati, Italy, 1984, pp. 3--12.
\newline\urlprefix\url{https://inis.iaea.org/collection/NCLCollectionStore/_Public/16/085/16085644.pdf}

\bibitem{Borovsky1987PAST-3}
I.~Borovsky, S.~Zhylkov, N.~Khyzhnyak, V.~Papkovich, To the theory of laser
  acceleration over dielectric comb, Problems of Atomic Science and Technics 34
  (1987) 69--70.

\bibitem{Borovsky1987PAST-4}
I.~Borovsky, S.~Zhylkov, N.~Khyzhnyak, V.~Papkovich, About the acceleration of
  the reb over a dielec-tric comb, Problems of Atomic Science and Technics 35
  (1987) 68--69.

\bibitem{Fernow1991BNL}
R.~Fernow,
  \href{https://inis.iaea.org/collection/NCLCollectionStore/_Public/23/008/23008350.pdf}{Acceleration
  using total internal reflection}, BNL Report No.52290 (1991) 1--17.
\newline\urlprefix\url{https://inis.iaea.org/collection/NCLCollectionStore/_Public/23/008/23008350.pdf}

\bibitem{Lenzner1998PRL}
M.~Lenzner, J.~Krüger, S.~Sartania, Z.~Cheng, C.~Spielmann, G.~Mourou,
  W.~Kautekand, F.~Krausz,
  \href{https://journals.aps.org/prl/pdf/10.1103/PhysRevLett.80.4076}{Femtosecond
  optical breakdown in dielectrics}, Phys. Rev. Lett. 80 (1998) 4076--4079.
\newblock \href {https://doi.org/10.1103/PhysRevLett.80.4076}
  {\path{doi:10.1103/PhysRevLett.80.4076}}.
\newline\urlprefix\url{https://journals.aps.org/prl/pdf/10.1103/PhysRevLett.80.4076}

\bibitem{Bendib2013LPB}
A.~Bendib, K.~Bendib-Kalache, C.~Deutsch, Optical breakdown threshold in fused
  silica with femtosecond laser pulses, Laser and Particle Beams 31~(3) (2013)
  523--529.
\newblock \href {https://doi.org/10.1017/S0263034613000396}
  {\path{doi:10.1017/S0263034613000396}}.

\bibitem{Soong2012AIP}
K.~Soong, R.~L. Byer, E.~R. Colby, R.~J. England, E.~A. Peralta,
  \href{https://doi.org/10.1063/1.4773749}{{Laser damage threshold measurements
  of optical materials for direct laser accelerators}}, AIP Conference
  Proceedings 1507~(1) (2012) 511--515.
\newblock \href
  {http://arxiv.org/abs/https://pubs.aip.org/aip/acp/article-pdf/1507/1/511/11481958/511\_1\_online.pdf}
  {\path{arXiv:https://pubs.aip.org/aip/acp/article-pdf/1507/1/511/11481958/511\_1\_online.pdf}},
  \href {https://doi.org/10.1063/1.4773749} {\path{doi:10.1063/1.4773749}}.
\newline\urlprefix\url{https://doi.org/10.1063/1.4773749}

\bibitem{Hobbs2012SPIE}
D.~Hobbs, B.~MacLeod, E.~Sabatino, Continued advancement of laser damage
  resistant optically functional microstructures, Proceedings of SPIE - The
  International Society for Optical Engineering 8530 (11 2012).
\newblock \href {https://doi.org/10.1117/12.976909}
  {\path{doi:10.1117/12.976909}}.

\bibitem{Peralta2013Nature}
E.~A. Peralta, E.~R.~J. Soong, K., E.~R. Colby, Z.~Wu, B.~Montazeri,
  C.~McGuinness, J.~McNeur, K.~J. Leedle, D.~Walz, E.~B. Sozer, B.~Cowan,
  B.~Schwartz, G.~Travish, R.~L. Byer,
  \href{https://doi.org/10.1038/nature12664}{Demonstration of electron
  acceleration in a laser-driven dielectric microstructure}, Nature 503 (2013)
  91--94.
\newblock \href {https://doi.org/10.1038/nature12664}
  {\path{doi:10.1038/nature12664}}.
\newline\urlprefix\url{https://doi.org/10.1038/nature12664}

\bibitem{Breur2013PRL}
J.~Breuer, P.~Hommelhoff,
  \href{https://link.aps.org/doi/10.1103/PhysRevLett.111.134803}{Laser-based
  acceleration of nonrelativistic electrons at a dielectric structure}, Phys.
  Rev. Lett. 111 (2013) 134803.
\newblock \href {https://doi.org/10.1103/PhysRevLett.111.134803}
  {\path{doi:10.1103/PhysRevLett.111.134803}}.
\newline\urlprefix\url{https://link.aps.org/doi/10.1103/PhysRevLett.111.134803}

\bibitem{England2014RMP}
R.~J. England, R.~J. Noble, K.~Bane, D.~H. Dowell, C.-K. Ng, J.~E. Spencer,
  S.~Tantawi, Z.~Wu, R.~L. Byer, E.~Peralta, K.~Soong, C.-M. Chang,
  B.~Montazeri, S.~J. Wolf, B.~Cowan, J.~Dawson, W.~Gai, P.~Hommelhoff, Y.-C.
  Huang, C.~Jing, C.~McGuinness, R.~B. Palmer, B.~Naranjo, J.~Rosenzweig,
  G.~Travish, A.~Mizrahi, L.~Schachter, C.~Sears, G.~R. Werner, R.~B. Yoder,
  \href{https://link.aps.org/doi/10.1103/RevModPhys.86.1337}{Dielectric laser
  accelerators}, Rev. Mod. Phys. 86 (2014) 1337--1389.
\newblock \href {https://doi.org/10.1103/RevModPhys.86.1337}
  {\path{doi:10.1103/RevModPhys.86.1337}}.
\newline\urlprefix\url{https://link.aps.org/doi/10.1103/RevModPhys.86.1337}

\bibitem{Wootton2016}
K.~P. Wootton, J.~McNeur, K.~J. Leedle,
  \href{https://doi.org/10.1142/S179362681630005X}{Dielectric laser
  accelerators: Designs, experiments, and applications}, Reviews of Accelerator
  Science and Technology 09 (2016) 105--126.
\newblock \href
  {http://arxiv.org/abs/https://doi.org/10.1142/S179362681630005X}
  {\path{arXiv:https://doi.org/10.1142/S179362681630005X}}, \href
  {https://doi.org/10.1142/S179362681630005X}
  {\path{doi:10.1142/S179362681630005X}}.
\newline\urlprefix\url{https://doi.org/10.1142/S179362681630005X}

\bibitem{Sun2021NJP}
L.~Sun, W.~Liu, J.~Zhou, Y.~Zhu, Z.~Yu, Y.~Liu, Q.~Jia, B.~Sun, H.~Xu,
  \href{https://dx.doi.org/10.1088/1367-2630/ac03cf}{{GV} m-1 on-chip particle
  accelerator driven by few-cycle femtosecond laser pulse}, New Journal of
  Physics 23~(6) (2021) 063031.
\newblock \href {https://doi.org/10.1088/1367-2630/ac03cf}
  {\path{doi:10.1088/1367-2630/ac03cf}}.
\newline\urlprefix\url{https://dx.doi.org/10.1088/1367-2630/ac03cf}

\bibitem{England_2022}
R.~England, U.~Niedermayer, L.~Schächter, T.~Hughes, P.~Musumeci, R.~Li,
  W.~Kimura,
  \href{https://dx.doi.org/10.1088/1748-0221/17/05/P05012}{Considerations for a
  {TeV} collider based on dielectric laser accelerators}, Journal of
  Instrumentation 17~(05) (2022) P05012.
\newblock \href {https://doi.org/10.1088/1748-0221/17/05/P05012}
  {\path{doi:10.1088/1748-0221/17/05/P05012}}.
\newline\urlprefix\url{https://dx.doi.org/10.1088/1748-0221/17/05/P05012}

\bibitem{Niedermayer_2022}
U.~Niedermayer, K.~Leedle, P.~Musumeci, S.~Schmid,
  \href{https://dx.doi.org/10.1088/1748-0221/17/05/P05014}{Beam dynamics in
  dielectric laser acceleration}, Journal of Instrumentation 17~(05) (2022)
  P05014.
\newblock \href {https://doi.org/10.1088/1748-0221/17/05/P05014}
  {\path{doi:10.1088/1748-0221/17/05/P05014}}.
\newline\urlprefix\url{https://dx.doi.org/10.1088/1748-0221/17/05/P05014}

\bibitem{Zhang2023TPS}
L.~Zhang, W.~Liu, H.~Xu, S.~Liu, Y.~Lu, Synchronous acceleration of
  subrelativistic particles in an inverse-cherenkov dielectric laser
  accelerator with tapered phase velocity, IEEE Transactions on Plasma Science
  51~(12) (2023) 3484--3491.
\newblock \href {https://doi.org/10.1109/TPS.2023.3338778}
  {\path{doi:10.1109/TPS.2023.3338778}}.

\bibitem{PhysRevAccelBeams.27.051303}
S.~Crisp, A.~Ody, J.~England, P.~Musumeci,
  \href{https://link.aps.org/doi/10.1103/PhysRevAccelBeams.27.051303}{Extended
  interaction length laser-driven acceleration in a tunable dielectric
  structure}, Phys. Rev. Accel. Beams 27 (2024) 051303.
\newblock \href {https://doi.org/10.1103/PhysRevAccelBeams.27.051303}
  {\path{doi:10.1103/PhysRevAccelBeams.27.051303}}.
\newline\urlprefix\url{https://link.aps.org/doi/10.1103/PhysRevAccelBeams.27.051303}

\bibitem{Palmer1980PA}
R.~B. Palmer, \href{http://cds.cern.ch/record/1107986}{{A Laser Driven Grating
  Linac}}, Part. Accel. 11 (1980) 81--90.
\newline\urlprefix\url{http://cds.cern.ch/record/1107986}

\bibitem{Zhang_2023}
L.~Zhang, W.~Liu, Y.~Liu, Q.~Jia, B.~Sun, H.~Xu, S.~Liu,
  \href{https://dx.doi.org/10.1088/1361-6463/acaaba}{Inverse cherenkov
  dielectric laser accelerator for ultra-relativistic particles}, Journal of
  Physics D: Applied Physics 56~(4) (2022) 045103.
\newblock \href {https://doi.org/10.1088/1361-6463/acaaba}
  {\path{doi:10.1088/1361-6463/acaaba}}.
\newline\urlprefix\url{https://dx.doi.org/10.1088/1361-6463/acaaba}

\bibitem{Vasiliev2023PAST}
A.~Vasiliev, A.~Povrozin, G.~Sotnikov,
  \href{https://doi.org/10.46813/2023-145-099}{General solution of the
  excitation problem of a symmetric flat dielectric structure by plane
  electromagnetic waves}, Problems of Atomic Science and Technics 145 (2023)
  99--102.
\newblock \href {https://doi.org/10.46813/2023-145-099}
  {\path{doi:10.46813/2023-145-099}}.
\newline\urlprefix\url{https://doi.org/10.46813/2023-145-099}

\bibitem{Breuer_2014_JPB}
J.~Breuer, J.~McNeur, P.~Hommelhoff,
  \href{https://dx.doi.org/10.1088/0953-4075/47/23/234004}{Dielectric laser
  acceleration of electrons in the vicinity of single and double grating
  structures—theory and simulations}, Journal of Physics B: Atomic, Molecular
  and Optical Physics 47~(23) (2014) 234004.
\newblock \href {https://doi.org/10.1088/0953-4075/47/23/234004}
  {\path{doi:10.1088/0953-4075/47/23/234004}}.
\newline\urlprefix\url{https://dx.doi.org/10.1088/0953-4075/47/23/234004}

\bibitem{CST-Particle}
\href{https://www.3ds.com/products/simulia/electromagnetic-simulation/particle-dynamics}{{CST
  Studio Suite. Particle Dynamics Simulation}}.
\newline\urlprefix\url{https://www.3ds.com/products/simulia/electromagnetic-simulation/particle-dynamics}

\bibitem{Vasiliev2018PAST}
A.~Vasiliev, A.~Dovbnya, A.~Yegorov, V.~Zaitsev, V.~Leshchenko, I.~Onischenko,
  A.~Povrozin, G.~Sotnikov,
  \href{https://vant.kipt.kharkov.ua/ARTICLE/VANT_2018_4/article_2018_4_289.pdf}{Works
  in the {NSC KIPT} on the creation and application of the {CPA} laser system},
  Problems of Atomic Science and Technology 116~(4) (2018) 289 – 293, cited
  by: 5.
\newline\urlprefix\url{https://vant.kipt.kharkov.ua/ARTICLE/VANT_2018_4/article_2018_4_289.pdf}

\bibitem{Breur2014PRSTAB}
J.~Breuer, R.~Graf, A.~Apolonski, P.~Hommelhoff,
  \href{https://link.aps.org/doi/10.1103/PhysRevSTAB.17.021301}{Dielectric
  laser acceleration of nonrelativistic electrons at a single fused silica
  grating structure: Experimental part}, Phys. Rev. ST Accel. Beams 17 (2014)
  021301.
\newblock \href {https://doi.org/10.1103/PhysRevSTAB.17.021301}
  {\path{doi:10.1103/PhysRevSTAB.17.021301}}.
\newline\urlprefix\url{https://link.aps.org/doi/10.1103/PhysRevSTAB.17.021301}

\bibitem{Plettner2006PRSTAB}
T.~Plettner, P.~P. Lu, R.~L. Byer,
  \href{https://link.aps.org/doi/10.1103/PhysRevSTAB.9.111301}{Proposed
  few-optical cycle laser-driven particle accelerator structure}, Phys. Rev. ST
  Accel. Beams 9 (2006) 111301.
\newblock \href {https://doi.org/10.1103/PhysRevSTAB.9.111301}
  {\path{doi:10.1103/PhysRevSTAB.9.111301}}.
\newline\urlprefix\url{https://link.aps.org/doi/10.1103/PhysRevSTAB.9.111301}

\bibitem{Peralta_PhD}
E.~A. Peralta, \href{https://purl.stanford.edu/ht547xt5560}{Accelerator on a
  chip: Design, fabrication, and demonstration of grating-based dielectric
  microstructures for laser-driven acceleration of electrons}, Ph.D. thesis,
  Stanford University, submitted to the Department of Applied Physics (March
  2015).
\newline\urlprefix\url{https://purl.stanford.edu/ht547xt5560}

\bibitem{Wei_PhD}
W.~Yelong, Investigations into dual-grating dielectric laser-driven
  accelerators, Ph.D. thesis, University of Liverpool, submitted to Faculty of
  Science and Engineering, School of Physical Sciences (August 2018).
\newblock \href {https://doi.org/10.17638/03022757}
  {\path{doi:10.17638/03022757}}.

\bibitem{Kozak:17}
M.~Koz\'{a}k, P.~Beck, H.~Deng, J.~McNeur, N.~Sch\"{o}nenberger, C.~Gaida,
  F.~Stutzki, M.~Gebhardt, J.~Limpert, A.~Ruehl, I.~Hartl, O.~Solgaard, J.~S.
  Harris, R.~L. Byer, P.~Hommelhoff,
  \href{https://opg.optica.org/oe/abstract.cfm?URI=oe-25-16-19195}{Acceleration
  of sub-relativistic electrons with an evanescent optical wave at a planar
  interface}, Opt. Express 25~(16) (2017) 19195--19204.
\newblock \href {https://doi.org/10.1364/OE.25.019195}
  {\path{doi:10.1364/OE.25.019195}}.
\newline\urlprefix\url{https://opg.optica.org/oe/abstract.cfm?URI=oe-25-16-19195}

\bibitem{Bolshov2021PAST}
O.~Bolshov, A.~Vasiliev, A.~Povrozin, G.~Sotnikov,
  \href{https://doi.org/10.46813/2021-136-057}{About the acceleration rate of
  relativistic beams by a surface wave in a dielectric laser accelerator},
  Problems of Atomic Science and Technics 136 (2021) 57--60.
\newblock \href {https://doi.org/10.46813/2021-136-057}
  {\path{doi:10.46813/2021-136-057}}.
\newline\urlprefix\url{https://doi.org/10.46813/2021-136-057}

\bibitem{Frandsen2006LP}
B.~R. Frandsen, S.~A. Glasgow, J.~B. Peatross,
  \href{https://doi.org/10.1134/S1054660X06090040}{Acceleration of free
  electrons in a symmetric evanescent wave}, Laser Physics 16 (2006)
  1311--1314.
\newblock \href {https://doi.org/10.1134/S1054660X06090040}
  {\path{doi:10.1134/S1054660X06090040}}.
\newline\urlprefix\url{https://doi.org/10.1134/S1054660X06090040}

\end{thebibliography}

\end{document}